\documentclass[12pt,a4paper]{iopart}
\usepackage{iopams}
\usepackage{epsfig}
\usepackage{subfigure}
\begin{document}
\title[\textsc{On the Mott formula for thermopower of QPC's}]{On the Mott formula for thermopower of non-interactions electrons in quantum point contacts}
\author{Anders Mathias Lunde\footnote[1]{lunan@fys.ku.dk} and Karsten Flensberg}
\address{Niels Bohr Institute, University of Copenhagen, DK-2100 Copenhagen, Denmark}
%
%-------------definition af newcommands-------------
\providecommand{\h}{\hslash}

\providecommand{\p}{\partial}

\providecommand{\eF}{\varepsilon_{\rm F}^{}}

\newcommand{\dd}{\textrm{d}} %bruges i intetraler og afledte.

\newcommand{\kb}{k_{\rm B}}

\newcommand{\om}{\omega}

\newcommand{\ve}{\varepsilon}
%---------------------------------------------------
%
\begin{abstract}
We calculate the linear response thermopower $S$ of a quantum
point contact using the Landauer formula and therefore assume
non-interacting electrons. The purpose of the paper, is to compare
analytically and numerically the linear thermopower $S$ of
non-interacting electrons to the low temperature approximation,
$S^{(1)}=(\pi^2/3e)\kb^2 T \p_{\mu}[\ln G(\mu,T=0)]$, and the
so-called Mott expression, $S^{\rm M}=(\pi^2/3e)\kb^2 T
\p_{\mu}[\ln G(\mu,T)]$, where $G(\mu,T)$ is the (temperature
dependent) conductance. This comparison is important, since the
Mott formula is often used to detect deviations from
single-particle behavior in the thermopower of a point contact.
\end{abstract}

\section{Introduction}

A narrow constriction in for example a two dimensional electron
gas makes a small cannel between two electron reservoirs. This
constriction is called a quantum point
contact\cite{QPC-let-review-Physics-today-Beenakker-1996}. The
width of the cannel can be controlled by a gate voltage and by
applying a small bias the phenomenon of quantized conductance as a
function of the width(i.e.~gate voltage) is observed at low
temperatures\cite{QPC-first-kvanticeret-conductance-Wees-1988}.
This quantization is due to the wave nature of the electronic
transport through the short ballistic point contact.
Experimentally\cite{Molenkamp-thermopower-QPC-first-PRL-1990,Molenkamp-thermocurrent-QPC-PRL-1992,Molenkamp-thermopower-review-SSAT-1992,Appleyard-Thermometer-QPC-thermopower-1998,Appleyard-many-body-effect-in-Thermopower-QPC-PRB-2000},
it is also possible to heat up one of the sides of the point
contact and thereby producing a temperature difference $\Delta T$
across the contact, which in turn gives an electric current (and a
heat current) though the point contact. By applying a bias $V$ in
the opposite direction of the temperature difference $\Delta T$
the two contributions to the electric current $I$ can be make to
cancel, which defines the thermopower $S$ as
\begin{equation}
S=\left. -\lim_{\Delta T \rightarrow 0}\frac{V}{\Delta
T}\right|_{I=0}. \label{eq:Thermopower-def}
\end{equation}
For a quantum point contact, the thermopower as a function of gate
voltage has a peak every time the conductance plateau changes from
one subband of the transverse quantization to the
next\cite{Molenkamp-thermopower-review-SSAT-1992,First-beregning-Thermopower-QPC-Streda-1989}.

In order to compare experiment and theory for the thermopower of a
point contact the so-called Mott formula,
\begin{equation}
S^{\rm M}\propto \p_{V_g}\left[\ln G(V_g,T)\right],
\label{eq:Cutler-Mott-approx-Vg}
\end{equation}
is often a valuable toll, because by differentiating the
experimentally found conductance $G(V_g,T)$ with respect to the
gate voltage $V_g$ one can see, if there is more information in
the thermopower that in the conductance. This additional
information could for example be many body
effects\cite{Appleyard-many-body-effect-in-Thermopower-QPC-PRB-2000},
since $S^{\rm M}$ is an approximation to the single-particle
thermopower. Note that this approximation is independent of the
specific form of the transmission $\mathcal{T}(\ve)$ though the
point contact. It is the purpose of this paper, to determine the
validity of the Mott approximation $S^{\rm M}$ and thereby decide
if it is really deviations from single-particle behavior the
experiments\cite{Appleyard-Thermometer-QPC-thermopower-1998,Appleyard-many-body-effect-in-Thermopower-QPC-PRB-2000,Thermopower-QPC-07-Preprint-James-2004}
observe or rather an artefact of this approximation.

\section{Thermopower from the Landauer formula}

For the sake of completeness, we begin by deriving the
single-particle thermopower formula in linear response to the
applied bias $V$ and temperature difference $\Delta T$. The
current though a ballistic point contact is found from the
Landauer formula\cite{Flensberg-Bruus-many-body-Book-2004}[p.111,
Eq.(7.30)]:
\begin{equation}
I=\frac{2e}{h}\int_0^{\infty}\!\!\!\!\dd \ve \mathcal{T}(\ve)
[f^0_L(\ve)-f^0_R(\ve)], \label{eq:Landauer-formula}
\end{equation}
where $\mathcal{T}(\ve)$ is the transmission and $f^0_i(\ve)$ is
the Fermi function for the right/left ($i=R,L$) lead. The Landauer
formula assumes non-interacting electrons and therefore so will
the derived thermopower formula. When a small bias
$V=(\mu_L-\mu_R)/(-e)$ and temperature difference $\Delta
T=T_L-T_R$ is applied, we can expand the distribution functions
around $\mu$, $T$ as ($|\Delta T|/T\ll 1$ and $|eV|\ll \mu$):
\begin{equation}
f^0_i(\ve)\simeq
f^0(\ve)-\p_{\ve}f^0(\ve)(\mu-\mu_i)-(\ve-\mu)\p_{\ve}f^0(\ve)\frac{T-T_i}{T},
\end{equation}
where $f^0(\ve)$ is the Fermi function with the equilibrium
chemical potential $\mu$ and temperature $T$ and $i=L,R$. To
obtain the thermopower eq.(\ref{eq:Thermopower-def}) we insert the
distribution functions in eq.(\ref{eq:Landauer-formula}) and set
it equal to zero and obtain:
\begin{eqnarray}
S(\mu,T)=\frac{1}{eT} \frac{\int_0^{\infty}\!\dd \ve
\mathcal{T}(\ve)(\ve-\mu)
[-\p_{\ve}f^0(\ve)]}{\int_0^{\infty}\!\dd \ve
\mathcal{T}(\ve)[-\p_{\ve}f^0(\ve)]},
\label{eq:exact-thermopower-single-particle-picture}
\end{eqnarray}
which is our exact single-particle formula.

\section{Approximations to the thermopower and there validity}

\subsection{The low temperature (first order)
approximation}

For $T=0$ we have $-\p_{\ve}f^0(\ve)=\delta(\ve-\mu)$, so the
numerator in
eq.(\ref{eq:exact-thermopower-single-particle-picture}) is zero,
i.e.~$S(\mu,T=0)=0$. For temperatures $\kb T$ much lower than the
scale of variation of $\mathcal{T}(\ve)$ and $\kb T \ll \mu$, we
can expand $\mathcal{T}(\ve)$ around $\mu$ to first order (i.e.~a
Sommerfeld expansion) to obtain:
\begin{equation}
\hspace{-2cm} S^{(1)}(\mu,T)=\frac{\pi^2}{3}\frac{k_{\rm
B}}{e}k_{\rm B} T \frac{1}{\mathcal{T}(\mu)}\frac{\p
\mathcal{T}(\mu)}{\p \ve}= %%
\frac{\pi^2}{3}\frac{\kb}{e}k_{\rm B} T \frac{1}{G(\mu,T=0)}
\frac{\p G(\mu,T=0)}{\p \mu},
\label{eq:thermopower-first-order-in-kT}
\end{equation}
where $G(\mu,T=0)$ is the conductance for zero temperature,
i.e.~$G(\mu,T=0)=\frac{2e^2}{h}\mathcal{T}(\mu)$.

\subsection{The Mott approximation and analytical considerations of its validity}

The Mott approximation\footnote[2]{In the early works by Mott and
co-workers
\cite{Mott-Book-1936,Cutler-Mott-formula-original-PR-1969} it was
actually the first order approximation
eq.(\ref{eq:thermopower-first-order-in-kT}), which was refereed to
as the Mott formula.}
\cite{Appleyard-Thermometer-QPC-thermopower-1998,Appleyard-many-body-effect-in-Thermopower-QPC-PRB-2000}
is
\begin{eqnarray}
S^{\rm M}(\mu,T)=\frac{\pi^2}{3}\frac{\kb}{e}k_{\rm B} T
\frac{1}{G(\mu,T)}\frac{\p G(\mu,T)}{\p \mu},
\label{eq:Cutler-Mott-approx}
\end{eqnarray}
where $G(\mu,T)$ is \emph{the temperature dependent} conductance
\begin{equation}
G(\mu,T)=\frac{2e^2}{h}\int_0^{\infty}\!\dd \ve
\mathcal{T}(\ve)[-\p_{\ve}f^0(\ve)].
\label{eq:conductace-any-temp}
\end{equation}
The form of $S^{\rm M}$ stated in
eq.(\ref{eq:Cutler-Mott-approx-Vg}) assumes that the chemical
potential and gate voltage are linear dependent. The Mott
approximation to the single-particle thermopower
eq.(\ref{eq:exact-thermopower-single-particle-picture}) and its
range of validity is not so obvious compared to the approximation
of the first order Sommerfeld expansion
eq.(\ref{eq:thermopower-first-order-in-kT}).

One way of comparing $S$ from
eq.(\ref{eq:exact-thermopower-single-particle-picture}) and
$S^{\rm M}$ is to differentiate eq.(\ref{eq:conductace-any-temp})
to obtain (assuming that $\mathcal{T}(\ve)$ is independent of
$\mu$):
\begin{eqnarray}
S^{\rm M}(\mu,T)= \frac{\pi^2}{3}\frac{\kb}{e}\frac{1}{G(\mu,T)}
\int_0^{\infty}\!\dd \ve \mathcal{T}(\ve)
\tanh\left(\frac{\ve-\mu}{2\kb T}\right) [-\p_{\ve}f^0(\ve)],
\label{eq:thermopower-CM-tanh-formel}
\end{eqnarray}
i.e.~by using the Mott formula we approximate $(\ve-\mu)/\kb T$ in
the integral by $(\pi^2/3)\tanh[(\ve-\mu)/(2\kb T)]$.

\begin{figure}
\begin{center}
\epsfig{file=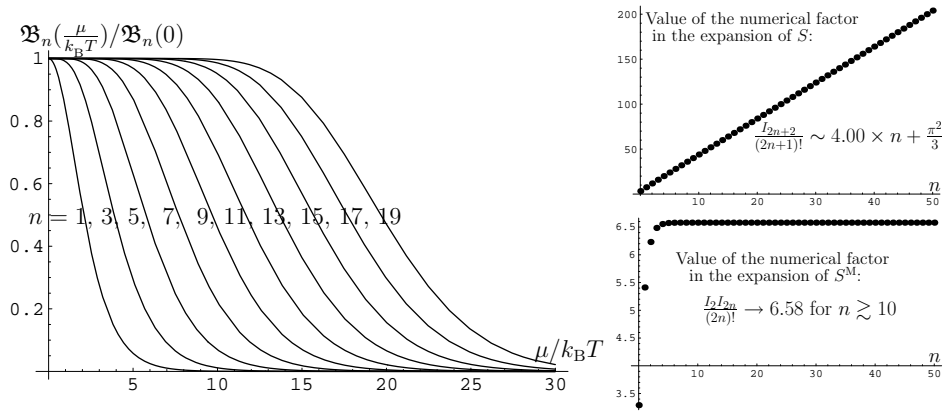,width=0.8\textwidth}
\caption{\footnotesize (Left): The approximation in
eq.(\ref{eq:integral-for-low-temperatures}) is pictured for odd
integer values of $n$ from 1(left) to 19(right) in
$\mathfrak{B}_n(\mu/\kb T)$. We note that $\mathfrak{B}_n(\mu/\kb
T)/\mathfrak{B}_n(0)\simeq 0$ for $\mu\gtrsim (10+n)\kb T$.
(Right): The numerical values of the factors in the series
expansions of the Mott approximation
eq.(\ref{eq:Cutler-Mott-series}) (lower)  and the exact linear
single-particle series expansion
eq.(\ref{eq:exact-thermopower-series} ) (upper).}
\label{fig:integral-approx}
\end{center}
\end{figure}

To compare $S$ and $S^{\rm M}$ in another way, we observe that for
low temperatures $\kb T\ll \mu$ the Mott approximation $S^{\rm M}$
simplifies to $S^{(1)}$
eq.(\ref{eq:thermopower-first-order-in-kT}), because
$G(\mu,T)\rightarrow\frac{2e^2}{h}\mathcal{T}(\mu)$ for
$T\rightarrow 0$, i.e. $S(\mu,T)=S^{(1)}(\mu,T)=S^{\rm M}(\mu,T)$
for $\kb T/\mu\rightarrow 0$. Therefore, we compare $S$ and
$S^{\rm M}$ by expanding both quantities in orders of $\kb T$ and
comparing order by order. Using
\begin{eqnarray}
\mathcal{T}(\ve)=\sum_{n=0}^{\infty}\frac{1}{n!}\frac{\p^n
\mathcal{T}(\mu)}{\p \ve^n}(\ve-\mu)^n,
\label{eq:expansion-of-Transmission}
\end{eqnarray}
we can exactly rewrite eq.(\ref{eq:conductace-any-temp})
\begin{eqnarray}
G&=& \frac{2e^2}{h} \sum_{n=0}^{\infty}\frac{1}{n!}\frac{\p^n
\mathcal{T}(\mu)}{\p \ve^n} \int_0^{\infty}\!\dd \ve (\ve-\mu)^n
[-\p_{\ve}f^0(\ve)]\nonumber\\
&=&\frac{2e^2}{h} \sum_{n=0}^{\infty}\frac{1}{n!}\frac{\p^n
\mathcal{T}(\mu)}{\p \ve^n} (\kb T)^n
\mathfrak{B}_n\Big(\frac{\mu}{\kb T}\Big),
\end{eqnarray}
where ($y=(\ve-\mu)/\kb T$)
\begin{eqnarray}
\hspace{-2.2cm} \mathfrak{B}_n\Big(\frac{\mu}{\kb T}\Big) \equiv
\int_{-\frac{\mu}{\kb T}}^{\infty}\!\!\!\!\dd y
\frac{y^n}{4\cosh^2(y/2)}
\quad \rightarrow \quad I_n \equiv
\int_{-\infty}^{\infty}\!\!\!\!\dd y \frac{y^n}{4\cosh^2(y/2)}
\quad \textrm{for}\ \kb T\ll\mu,
\label{eq:integral-for-low-temperatures}
\end{eqnarray}
where we note that $I_{2n+1}=0$ for all integer $n$. Numerically
it turns out, that $\mathfrak{B}_n(\mu/\kb
T)/\mathfrak{B}_n(0)\simeq 0$ for $\mu\gtrsim (10+n)\kb T$ as seen
in figure \ref{fig:integral-approx}. The integral $I_n$ can be
calculated and the first values are:
\begin{equation}
I_0=1,\ I_2=\frac{\pi^2}{3},\ I_4=\frac{7\pi^4}{15},\
I_6=\frac{31\pi^6}{21},\ I_8=\frac{127\pi^8}{15},\ \ldots
\end{equation}
Using the approximation
eq.(\ref{eq:integral-for-low-temperatures}) we get
\begin{eqnarray}
G(\mu,T)\simeq\frac{2e^2}{h}
\sum_{n=0}^{\infty}\frac{1}{(2n)!}\frac{\p^{2n}
\mathcal{T}(\mu)}{\p \ve^{2n}} I_{2n} (\kb T)^{2n}.
\end{eqnarray}
This leads to a Mott approximation to the thermopower for low
temperatures as
\begin{eqnarray}
S^{\rm M}(\mu,T)\simeq\frac{\kb}{e} \frac{1}{G(\mu,T)}
\frac{2e^2}{h} \left[ \sum_{n=0}^{\infty}\frac{I_2
I_{2n}}{(2n)!}\frac{\p^{2n+1} \mathcal{T}(\mu)}{\p \ve^{2n+1}}
(\kb T)^{2n+1}\right].
\label{eq:Cutler-Mott-series}
\end{eqnarray}
Writing the exact single-particle thermopower $S$
eq.(\ref{eq:exact-thermopower-single-particle-picture}) by using
eq.(\ref{eq:expansion-of-Transmission}) and the approximation of
low temperatures eq.(\ref{eq:integral-for-low-temperatures}), we
get
\begin{eqnarray}
S(\mu,T)\simeq\frac{\kb}{e} \frac{1}{G(\mu,T)}
\frac{2e^2}{h} \left[
\sum_{n=0}^{\infty}\frac{I_{2n+2}}{(2n+1)!}\frac{\p^{2n+1}
\mathcal{T}(\mu)}{\p \ve^{2n+1}} (\kb T)^{2n+1}\right].
\label{eq:exact-thermopower-series}
\end{eqnarray}
We see that both formulas only have odd term in $\kb T$ and the
first order term is the same (which is $S^{(1)}$). However, none
of the higher order terms are the same and on figure
\ref{fig:integral-approx}(right) the different numerical factors
of the two series expansions are seen to behave very differently
as the power of $\kb T$ grows:
\begin{eqnarray}
\frac{I_{2n+2}}{(2n+1)!}\sim 4.00\times n +\frac{\pi^2}{3} \quad
\textrm{and} \quad \frac{I_2I_{2n}}{(2n)!}\rightarrow 6.58 \ \
\textrm{for} \ n\gtrsim 10.
\end{eqnarray}

So the Mott approximation is better the smaller the temperature
compared to $\mu$, but not a bad approximation for moderate
temperatures (i.e.~$\kb T$ comparable to other energy scales) as
we shall see numerically. Note that if the approximation
eq.(\ref{eq:integral-for-low-temperatures}) is not valid, then we
have all powers of $\kb T$.

\section{Comparison of the approximations to the exact single-particle thermopower from numerical integration}

We need a specific model for the transmission to do a numerical
comparison of $S$ from
eq.(\ref{eq:exact-thermopower-single-particle-picture}) to $S^{\rm
M}$ and $S^{(1)}$. Using a harmonic potential in the point
contact, i.e.~a saddle point potential, a transmission in the form
of a Fermi function can be
derived\cite{Buttiker-transmission-saddle-point-potential-PRB-1990}:
\begin{eqnarray}
\mathcal{T}(\ve)=\sum_{n=1}^{n_{\rm max}}
\frac{1}{\exp(\frac{n\ve_0-\ve}{\ve_s})+1},
\label{eq:Transmission-Buttiker}
\end{eqnarray}
where $\ve_s$ is the smearing of the transmission between the
steps and $\ve_0$ is the length of the steps (often called subband
spacing). In terms of the harmonic potential
$V(x,y)=\textrm{const}-m\om_x^2x^2/2+m\om_y^2y^2/2$, where $x$ is
along the cannel, we have $\ve_0=\h \omega_y$ and $\ve_s=\h
\om_x/(2\pi)$. Other functional forms of $\mathcal{T}$ have also
been tested, but as along as they have the same graphical
structure (such as for example a $\tanh$ dependence) the same
conclusions are obtained.

\begin{figure}
\begin{center}
\epsfig{file=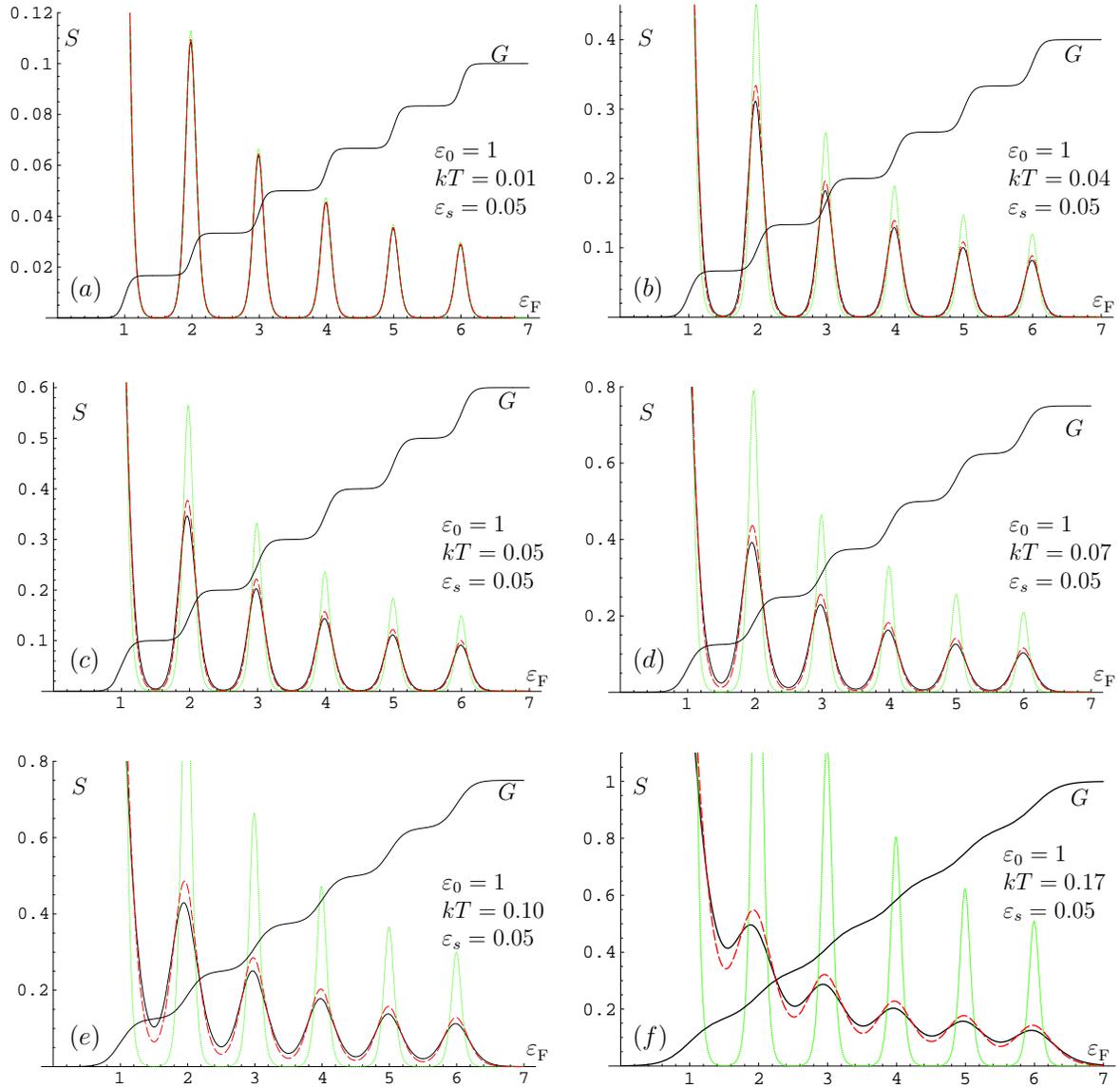,width=1.0\textwidth}
\caption{\footnotesize Thermopower $S$ from numerical integration
of eq.(\ref{eq:exact-thermopower-single-particle-picture}) (black
solid line), the Mott formula $S^{\rm M}$
eq.(\ref{eq:Cutler-Mott-approx}) (red dashed line) and the first
order approximation $S^{(1)}$
eq.(\ref{eq:thermopower-first-order-in-kT}) (green dotted line).
From figure (a) to (f) the temperature is turned from the low
temperature regime $\kb T < \ve_s$ to $\kb T>\ve_s$ in small
steps. The smearing of the transmission $\ve_s$ is keep constant
and note that $\ve_s, \kb T\ll \ve_0$ and $\ve_s, \kb T\ll \eF$ in
all the graphs. The thermopowers are all in units of $\kb/e$, but
note the different magnitudes of the thermopower from (a) to (f).
The conductance $G$ is shown (in arbitrary units) for comparison.}
\label{fig:Thermopower}
\end{center}
\end{figure}

Three regimes of temperatures relevant to experiments are
investigated numerically:
\begin{equation}
\hspace{-1.5cm} \kb T<\ve_s \ (\textrm{fig.\ref{fig:Thermopower}
(a)}), \quad \kb T\sim \ve_s \ (\textrm{fig.\ref{fig:Thermopower}
(b-d)})\quad \textrm{and}\quad \kb T>\ve_s \
(\textrm{fig.\ref{fig:Thermopower} (e-f)}).
\label{eq:regimes-of-temp}
\end{equation}
The thermopower $S$ for the transmission model
eq.(\ref{eq:Transmission-Buttiker}) is found from numerical
integration of
eq.(\ref{eq:exact-thermopower-single-particle-picture}) and
compared to the Mott approximation $S^{\rm M}$
eq.(\ref{eq:Cutler-Mott-approx}) and the first order approximation
$S^{(1)}$ eq.(\ref{eq:thermopower-first-order-in-kT}). In all
three regimes, we have a staircase conductance, so $\kb T\ll
\ve_0$, and $G(\mu,T)$ is also shown in the figures (in arbitrary
units) for comparison. Furthermore, $\mu=\eF$ is of order $\ve_0$,
so the approximation $\kb T\ll \eF$ used for example in
eq.(\ref{eq:integral-for-low-temperatures}) is indeed very good.
Note that all energies in the figures are given in units of the
step length $\ve_0$.

The information obtained from the numerical calculations is the
following. Figure~\ref{fig:Thermopower}(a-b) shows that for $\kb
T$ being the lowest energy scale both approximations work very
well as expected from the analytical considerations. When the
temperature becomes comparable to the smearing of the steps, $\kb
T\sim \ve_s$, the Culter-Mott formula works well and better than
the first order approximation as seen in
fig.~\ref{fig:Thermopower}(b-d). For $\kb T$ bigger than $\ve_s$
the Mott approximation still works quit well whereas $S^{(1)}$ is
not a good approximation anymore. The reason for the Mott
approximation to work well is found in the similar terms in the
analytic temperature expansions eq.(\ref{eq:Cutler-Mott-series})
and (\ref{eq:exact-thermopower-series}). Note that as $\kb T$
increases both $S^{(1)}$ and $S^{\rm M}$ show a tendency to
overestimate $S$ at the peaks and underestimate it at the valleys.

In summary, we have found that the Mott approximation to the
single-particle thermopower is a fairly good approximation as
along as the temperature is smaller than the Fermi level, but $\kb
T$ can be both compatible and larger than the smearing of the
transmission $\ve_s$. However, to rule out any doubt one could use
an experimental determination of $\mathcal{T}(\ve)$ from the (very
low temperature) conductance to find the single-particle
thermopower from
eq.(\ref{eq:exact-thermopower-single-particle-picture}), which
could perhaps give an interesting comparison to the experimental
result. Thereby one would obtain an even more convincing statement
of deviations from single-particle behavior in the thermopower.

\section{Acknowledgements}

We will like to thank James T. Nicholls for sharing his
experimental results with us and for discussions of the
thermopower in point contacts in general.

\end{document}